\documentclass[%
 reprint,
 amsmath,amssymb,
 aps,twocolumn,prl
]{revtex4-1}

\usepackage{graphicx}
\usepackage{hyperref}
\usepackage{multirow}
\usepackage{amsfonts}

\newcommand{\eqn}[1]{\begin{eqnarray} #1 \end{eqnarray}}
\newcommand{\tit}[1]{\textit{#1}}
\newcommand{\tbf}[1]{\textbf{#1}}
\newcommand{\trm}[1]{\textrm{#1}}
\newcommand{\doo}[1]{#1^{\textrm{\textbf{do}}}}

\newcommand{\tr}[1]{  \textrm{Tr}\left[ #1 \right]  }

\newcommand{\zum}[2]{\displaystyle\sum_{#1}^{#2}}

\begin{document}

\title{Quantum causal models via QBism: the short version}

\author{Jacques Pienaar}
\affiliation{
 International Institute of Physics, Universidade Federal do Rio Grande do Norte, Campus Universitario, Lagoa Nova, Natal-RN 59078-970, Brazil.
}

\date{\today}



\begin{abstract}
This paper is a concise summary of the main ideas presented in the companion paper, arXiv:1806.00895 \cite{JACQUES}. I present the proposed definition of a quantum causal model with minimal background and justification, focusing only on its essential physical properties. The mathematical structure and definitions are provided as an Appendix. I discuss the possible physical significance of the fact that the model is symmetric under causal inversion.
\end{abstract}

\maketitle

\section{The main results in a nutshell}

For macroscopic (hence ``classical") systems such as the human body, economic markets, social organizations and factories, there exist a variety of causal models which -- their differences notwithstanding -- agree on the essential features of causality and on the methods for inferring causal relations from statistics \cite{PEARL,SGS,WOODW}.  Roughly, a \tit{causal model} refers to a pair $\{P(\tbf{X}),G(\tbf{X}) \}$, where $P(\tbf{X})$ is a probability distribution for a set of random variables $\tbf{X}$, and $G(\tbf{X})$ is a directed acyclic graph (DAG) with nodes corresponding to members of $\tbf{X}$, together with (i) a set of constraints on the pair $\{P(\tbf{X}),G(\tbf{X}) \}$ that must be satisfied in order for $G(\tbf{X})$ to be a \tit{possible explanation} of $P(\tbf{X})$ (called the \tit{Markov condition}), and (ii) a set of rules that allow us to obtain a new causal model $\{P(\tbf{X}|\, \trm{\tbf{action}}),G(\tbf{X}|\trm{ \tbf{action}}) \}$ which describes what happens when a certain \tbf{action} is performed (in classical models, the \tbf{action} is usually an \tit{intervention}). (Note: precise definitions of italicized terms will be given in the Appendix).

Recent efforts to extend this kind of causal modeling to microscopic quantum systems have resulted in a variety of approaches, which so far seem to have little in common with one another \cite{ALLEN,PIEBRUK,FRITZ,COSHRAP,HLP,POLLOCKPRA,POLLOCKPRL,CDP,SELBY} . The model presented here and in Ref. \cite{JACQUES} adds to this confusion by proposing yet another approach, having certain novel features. Its main novelty is symmetry under \tit{causal inversion} (reversal of the directions of all causal arrows). To some extent this might be expected, since the fundamental laws governing microscopic systems are believed to be time-reversible. However, this expectation hinges on the controversial question of what exactly is the relationship between causality and the direction of time \cite{PRICEBOOK}. The present work contributes to this discussion by arguing that symmetry under causal inversion is possible in an explicit model. To make sense of this counter-intuitive claim, it is proposed that the direction of causality is observer-dependent.

The motivation for the model in Ref. \cite{JACQUES} is summarized in the following section. The proposed model differs from a classical causal model in two essential ways. First, the proposed \tit{Quantum Markov Condition} rejects the classical assumption that variables correlated only by virtue of their common causes should factorize when one conditions on those causes (the so-called quantitative part of Reichenbach's common-cause principle \cite{CAVLAL}); in this respect the model is consistent with other quantum causal models. Secondly, the \tbf{actions} considered in the quantum causal model include, in addition to \tit{interventions}, the possibility to \tit{not} perform some measurement that was originally present in the model, which I call an \tit{un-measurement}. Classically, it is customary to assume that measurements are non-disturbing, in which case the result of an \tit{un-measurement} is trivially equivalent to ignoring the outcome of the measurement. Quantum \tit{un-measurements}, by contrast, are non-trivial and are therefore better regarded as a special type of \tbf{action}, though distinct from an \tit{intervention}. This inclusion of \tit{un-measurements} among the allowed \tbf{action}s is a completely novel feature.

In order to compute the probabilities for \tit{un-measurements}, we refer to the work of the Quantum Bayesians (QBists). In seeking to reformulate quantum theory as a theory of rational inference, the QBists have already derived precisely the equation we need to describe \tit{un-measurements}, using the concept of a \tit{symmetric informationally-complete} (SIC) measurement \cite{QBCOH,QPLEX}. The present work may therefore be understood as the result of applying the QBist program to the task of causal modeling.

These simple ingredients lead to a model that is symmetric under causal inversion, which means that if $P(\tbf{X})$ satisfies the \tit{Quantum Markov Condition} for a DAG $G(\tbf{X})$, then it is also satisfied for the causally inverted DAG $G^*(\tbf{X})$, obtained by reversing the direction of all causal arrows in $G(\tbf{X})$. It is argued that this symmetry might have a physical interpretation, as allowing for \tit{interventions} by hypothetical observers whose `causal arrow' runs contrary to our own.

\section{Motivation for the approach}

The quantum causal model presented here is motivated by the following observation. Classically, the pair $\{ P(\tbf{X}),G(\tbf{X}) \}$ is entirely sufficient to deduce how the system will respond to interventions. Furthermore, the \tit{Markov Condition} is sufficient to derive inequality constraints on $P(\tbf{X})$ imposed by the causal structure $G(\tbf{X})$ (of which the famous Bell inequalities are a special case), which must be satisfied if the causal model is to qualify as a valid description of the system. Since quantum systems routinely violate these inequalities in experiments where the causal structure is thought to be known (and if we do not wish to concede the existence of causal influences hidden from us) a natural recourse is to define a new kind of causal model that can accommodate such violations. 

In the literature, the most common solution is to replace the simple DAG $G(\tbf{X})$ with a richer structure, such as a quantum process theory, a generalized quantum network (quantum comb), or something similar. This new structure provides the causal relations, indicates the results of interventions and implies the known bounds on quantum correlations (viz. the amount by which they violate classical inequalities). A potential weakness with this broad approach is that there are many different ways to do it, as shown by the diverse literature \cite{ALLEN,COSHRAP,POLLOCKPRA,POLLOCKPRL,FRITZ,PIEBRUK,HLP,ORESH,RIED,SELBY,RIEDPHD}. 

The approach taken here is grounded on the premise that a quantum causal model, like its classical counterpart, does not require anything more than a DAG $G(\tbf{X})$ to represent causal structure. The particular structure of quantum probabilities must therefore be located within an appropriately defined \tit{Quantum Markov Condition}, and in the rules that describe how the system responds to certain \tbf{actions}. This idea turns out to be remarkably restrictive, and almost entirely determines the properties of the resulting model.

One obstacle to this program is that, as argued in Ref. \cite{JACQUES} and supported by the literature, any reasonable choice of \tit{Quantum Markov Condition} is no longer sufficient to determine the bounds on quantum correlations, at least without supplementing $G(\tbf{X})$ directly with additional structure, which would be contrary to our aim. To get around this, it is observed that the constraints on the allowed quantum probabilities can be re-cast as constraints on the way in which a quantum system responds to an \tit{un-measurement}, that is, the rule that tells us what probabilities to expect when a certain measurement is \tit{not} performed, in terms of the probabilities that obtain when the measurement is present.

This rule, known to the Quantum Bayesians as the \tit{Urgleichung} \cite{QBCOH}, implies certain constraints on the space of allowed probabilities which enable bounds on quantum correlations to be derived from it \cite{QPLEX}. Since the rule is defined under the assumption that the \tit{un-measurement} in question is a \tit{SIC-measurement}, we are thus led to assume that an idealized experiment (for the purposes of causal inference) is one in which all measurements are SIC-measurements. This rather strong assumption determines the various properties of the \tit{Quantum Markov Condition} proposed in Ref. \cite{JACQUES}, including the property that the quantitative part of Reichenbach's Principle no longer holds in general.

Another obstacle to the program is that, unlike the classical case, the pair $\{ P(\tbf{X}),G(\tbf{X}) \}$ is no longer sufficient in general to deduce the statistics that result from an \tit{intervention}. However, it turns out to still be possible provided that the DAG belongs to a special sub-class of graphs called \tit{QDAG}s, in which the variables can be partitioned into a set of causally ordered `slices' \cite{JACQUES}. By requiring $G(\tbf{X})$ to be a \tit{QDAG}, this obstacle is overcome.

The end result of these considerations is that it is possible to define a quantum causal model capable of making predictions about both \tit{interventions} and \tit{un-measurements} based only on the pair $\{ P(\tbf{X}),G(\tbf{X}) \}$ and the \tit{Quantum Markov Condition}; moreover the model goes some way towards imposing bounds on quantum correlations (but see Ref. \cite{QPLEX} and the remarks in Sec. \tbf{IV.A} of Ref. \cite{JACQUES} for qualifications of this claim). Although much of the structure of the model is derived by first considering its representation as a quantum circuit, the key features of the model refer only to relations between conditional probability distributions, and therefore do not depend on the details of the circuit. Thus, unlike other models, one can remain ambivalent about whether or not the circuit represents an objective `underlying' structure or merely a convenient fiction; the only details relevant for causal inference are the observed statistics and the raw causal relations $G(\tbf{X})$.

\section{Key properties of the model}

\begin{table*}[ht] 
\caption{Key properties of the Quantum Markov Condition \label{fig:table}}
\centering
\begin{tabular}{p{0.1\linewidth}|p{0.15\linewidth}p{0.15\linewidth}|p{0.15\linewidth}p{0.15\linewidth}|}
\cline{2-5}
 & \multicolumn{2}{ c| }{\tbf{Classical}}  & \multicolumn{2}{ c| }{\tbf{Quantum}} \\ \cline{2-5}
 & \multicolumn{1}{ c| }{Common Cause} & \multicolumn{1}{ c| }{Common Effect} & \multicolumn{1}{ c| }{Common Cause} & \multicolumn{1}{ c| }{Common Effect} \\ \cline{1-5}
\multicolumn{1}{ |c| }{A,B correlated?} & \multicolumn{1}{ c| }{\checkmark} & \multicolumn{1}{ c| }{$\times$} & \multicolumn{1}{ c| }{$\times$} & \multicolumn{1}{ c| }{$\times$}      \\ \cline{1-5}
\multicolumn{1}{ |c| }{A,B correlated conditional on C?} & \multicolumn{1}{ c| }{$\times$} & \multicolumn{1}{ c| }{\checkmark} & \multicolumn{1}{ c| }{\checkmark} & \multicolumn{1}{ c| }{\checkmark}      \\ \cline{1-5}
\end{tabular}
\end{table*}

\begin{figure}[!htb]
\centering\includegraphics[width=0.8\linewidth]{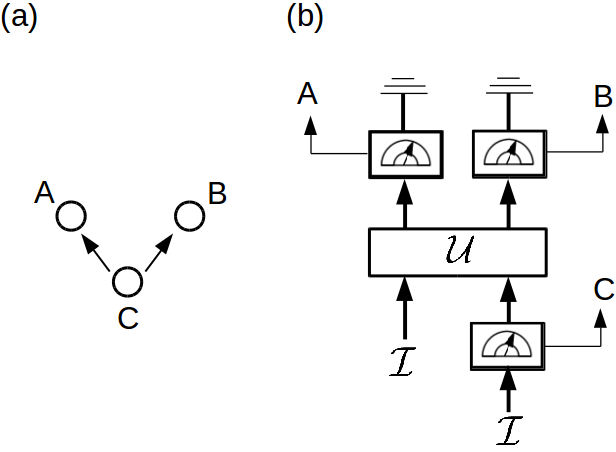}
\caption{(a) A DAG representing a common cause and (b) one of the possible functional models corresponding to this DAG. The meter-boxes represent SIC-measurements, $\mathcal{U}$ is a unitary gate and $\mathcal{I}$ represents the maximally mixed state. Bold wires represent quantum systems, and the thin wires exiting from the side of the meter boxes represent classical systems (i.e. measurement data). The quantum systems are discarded at the end of the process, but the classical records are retained.}
\label{fig:commcause}
\end{figure}

\begin{figure}[!htb]
\centering\includegraphics[width=0.8\linewidth]{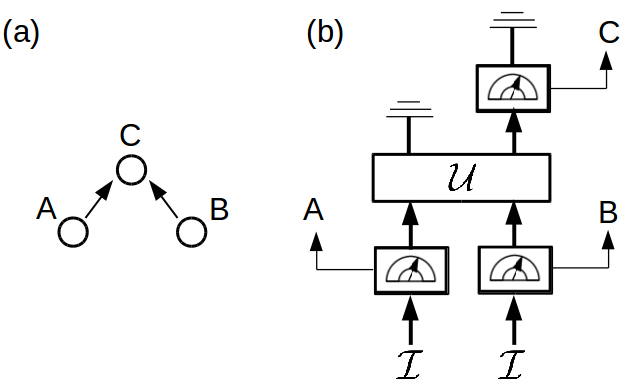}
\caption{(a) A DAG representing a common effect and (b) one of the possible functional models corresponding to this DAG.}
\label{fig:commeffect}
\end{figure}

The properties of the quantum causal model can be summarized by considering just two paradigmatic cases: the common cause and the common effect (Figs. \ref{fig:commcause},\ref{fig:commeffect}). The main features are summarized in Table \ref{fig:table}. As is evident from the table, the classical constraints on the correlations among the variables differ between the common cause and common effect, indicating a causal asymmetry. By contrast, the quantum constraints on correlations are exactly the same for the common cause and the common effect, which is due to the inherent symmetry of the Quantum Markov Condition. These features are explained below.

Fig. \ref{fig:commcause} (a) shows a DAG $G(A,B,C)$ representing a common cause, and Fig. \ref{fig:commcause} (b) shows a possible functional model (quantum circuit) for this DAG. Let $\{ P(A,B,C), G(A,B,C) \}$ be a quantum causal model for the DAG of Fig. \ref{fig:commcause} (a). If this were a \tit{classical} causal model, then the following three constraints would be expected to hold (from the Classical Markov Condition \cite{PEARL,SGS} ):\\ 
(i) $A$ and $B$ should generally be correlated if one does not condition on $C$;\\
(ii) The correlation between $A$ and $B$ should disappear when one conditions on $C$.\\
The constraint (ii) is called the \tit{quantitative part of Reichenbach's Principle of Common Causes}. The constraint (i) has been called the \tit{Principle of Independence} \cite{PRICEBOOK}, because it expresses the commonplace intuition that variables are independent prior to interaction, and not afterwards. However, this is something of a misnomer, since the important part is the `\tit{...and not afterwards}', which implies that variables with a shared history tend to be correlated; it would therefore be better named the `Principle of Correlation'. Note that (i) and (ii) should not be confused with the logically independent \tit{qualitative part of Reichenbach's Principle of Common Causes}, which may be stated as:\\
(iii) If two variables are correlated (and neither is a cause of the other) they must share a common cause. \\
This constraint (iii) is equivalent to saying that the variables must be independent if they do not share a common cause, but note that this would still permit a model in which $A$ and $B$ were required to be independent despite having a common cause $C$, and hence (iii) does not have the power to imply (i). All three constraints (i),(ii),(iii) generally hold in classical causal models. 

Regarding Fig. \ref{fig:commcause}, it is apparent that (ii) will be false for our quantum causal model (i.e. according to our Quantum Markov Condition). It is generally possible to arrange for $A,B$ to become independent conditional on a \tit{particular} value of $C$, for instance by arranging that the post-measurement state $\Pi_c$ conditioned on $C=c'$ is a product state between the Hilbert spaces subsequently measured at $A,B$ in the circuit of Fig. \ref{fig:commcause} (b). However, since the model assumes that the set $\{ \frac{1}{d_C} \Pi_c : \forall c \in C \}$ is a SIC-POVM, the remaining values $c \neq c'$ cannot also correspond to projectors with such a product form. Hence, simply conditioning on the outcome of $C$ will typically not eliminate correlations between $A,B$, and so (ii) must be rejected as a constraint. 

Actually, the situation is even more drastic: the correlations between $A,B$ \tit{only exist} when one conditions on the outcome of $C$. If one does not condition on $C$, then the joint quantum state being measured by $A,B$ is maximally mixed and thus uncorrelated. Hence the constraint (i) is also false in our quantum causal model. Note that this is purely due to the fact that SIC-measurements belong to a special class of quantum processes that are \tit{unbiased}, meaning that the mere fact that $C$ was performed cannot change the maximally mixed state into some other state, if one does not also condition on the outcome $C=c$. One might perhaps object to the model's assumption that the quantum state \tit{begins} in the maximally mixed state, as is evident from the circuit Fig. \ref{fig:commcause} (b) and is indeed part of the definition of the \tit{functional model}. This objection fails because, upon reflection, the only occasions we have to know the state of a system are those in which we prepared it in that state by first performing some measurement upon it and conditioning on the outcome of that measurement. Prior to the preparation procedure, the state is completely unknown and therefore best described as maximally mixed. If this aspect of the causal model appears strange, it is only because we are accustomed to leaving the preparation procedure implicit in the definition of initial state, rather than including it as a process explicitly within the model, as it is here.

Turning now to the common effect scenario of Fig. \ref{fig:commeffect}, note that the following constraints on $P(A,B,C)$ would be expected to hold for a \tit{classical} causal model:\\
(iv) $A$ and $B$ should generally be independent if one does not condition on $C$;\\
(v) $A$ and $B$ should generally be correlated if one conditions on $C$.\\
The constraint (iv) is just a special instance of (iii), the \tit{qualitative part of Reichenbach's Principle of Common Causes}, when $A,B$ have no common cause. To see that this is also true in the quantum causal model, note that $A,B$ represent measurements on independently prepared subsystems in the functional model Fig. \ref{fig:commeffect} (b). The constraint (v) is widely known as \tit{Berkson's effect} in statistics. It remains true for our quantum causal model, since post-selecting on an outcome $C=c$ corresponds to post-selecting the final state of the circuit Fig. \ref{fig:commeffect} (b) to be the pure state $\Pi_c$, and this will typically induce correlations between $A,B$. Note that these post-selection-induced correlations can violate the classical bound (i.e. Bell inequalities) precisely up to the quantum bound; this `quantum Berkson effect' has been studied in Ref. \cite{SPEKBERK}.

In summary, the proposed model rejects the \tit{Principle of Independence} and the \tit{quantitative} part of Reichenbach's Principle of Common Causes, (i),(ii), while upholding the \tit{qualitative} part of Reichenbach's Principle and Berkson's effect, (iii),(iv),(v). While the rejection of (ii) is relatively commonplace in the literature, the rejection of (i) is somewhat more of a novelty. Note that the model's symmetry under causal inversion depends on this rejection, since otherwise one could use (i) to differentiate between the common cause and the common effect by checking for the existence of a-priori correlations between $A$ and $B$. 

\begin{figure*}[p!]
\centering\includegraphics[width=15cm]{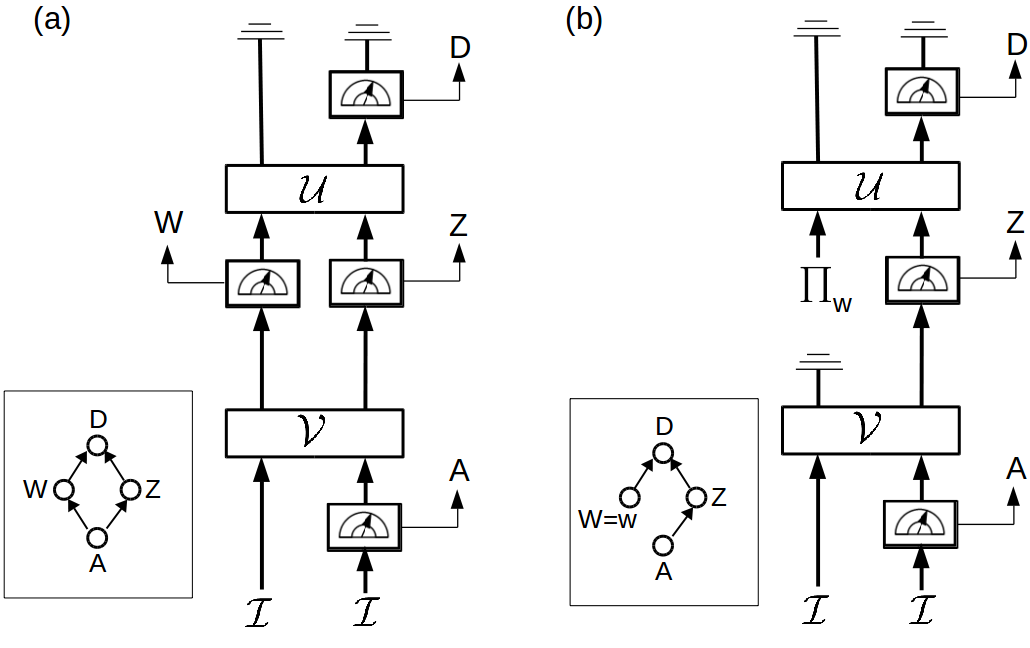}
\caption{(a) One of the possible functional models corresponding to a QDAG with four nodes $A,W,Z,D$ (shown inset). The meter-boxes represent SIC-measurements, $\mathcal{U},\mathcal{V}$ are unitary gates and $\mathcal{I}$ represents the maximally mixed state. (b) The functional model resulting from an intervention on $W$ to set $W=w$, and the corresponding post-intervention QDAG (shown inset).}
\label{fig:intervene}
\end{figure*}

\begin{figure*}[p!]
\centering\includegraphics[width=15cm]{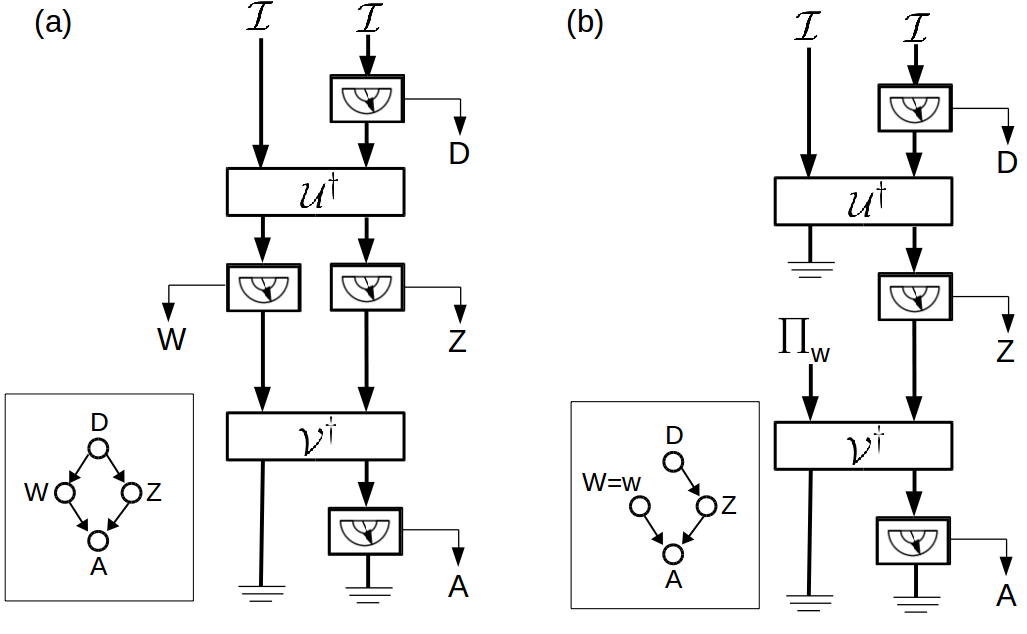}
\caption{(a) A possible functional model corresponding to the causal inverse QDAG (shown inset), obtained by time-reversing the circuit of Fig. \ref{fig:intervene} (a) and replacing $\mathcal{U},\mathcal{V}$ with $\mathcal{U}^{\dagger},\mathcal{V}^{\dagger}$. The result of an intervention on the reversed circuit (and its corresponding QDAG) is shown in (b).}
\label{fig:reverse}
\end{figure*}

\section{Interpretation of the causal symmetry}

At first glance, the symmetry of the quantum causal model under causal inversion seems to be a weakness. Imagine, for instance, that you are given the statistical data $P(A,B,C)$ and asked to decide, based only upon this data, whether the variable $C$ is a common cause of $A,B$ as in Fig. \ref{fig:commcause} (a), or a common effect as in Fig. \ref{fig:commeffect} (a). Due to the symmetry of the model, one cannot tell the difference without performing an intervention. Yet, the difference would become obvious after performing an intervention on $C$: if it is a common cause, then $A,B$ would each remain correlated to $C$ after the intervention, whereas if it is a common effect, these correlations would disappear. In other words, the common cause should allow $C$ to \tit{signal} to $A$ and $B$, but the common effect would not.

On this basis, one might argue that the constraints specified by the Quantum Markov Condition are too weak to do effective causal inference. Given that there is a preferred direction of causality within the system (eg. common cause versus common effect), then it seems reasonable that we should at least sometimes be able to infer this direction from the observed statistics $P(A,B,C)$ prior to any interventions. Yet, the proposed Quantum Markov Condition renders such causal inference impossible without interventions. It therefore seems to be ignoring useful causal information.

On further inspection, the matter is more subtle. The symmetry of the statistics $P(A,B,C)$ is enforced by the property of SIC-measurements that they are \tit{informationally-symmetric}, which means that the pre-measurement state of the system conditioned on the measurement outcome is the same as the post-measurement state conditioned on the outcome (this concept was recently introduced in the context of quantum causal inference \cite{RIED}). This means that in order to fully specify the measurement, it is not necessary to make any formal distinction between the ``input" and the ``output" of the measurement. Consequently, such measurements are agnostic about the overall direction of causality: one can switch the labels of input/output and still obtain a valid SIC-measurement. If we were to admit into our causal model measurements that were not \tit{informationally-symmetric}, and thus which might enable us to infer the overall causal direction without performing an intervention, we would have to specify \tit{a priori} a preferred labeling of input/output and thereby impose on the system the very causal arrow that we would seek to ``infer" from the measurement statistics!

Still, one might insist that there exists an intrinsic direction of causality in the system, whether or not it can be inferred from the data prior to interventions. After all, when we finally do perform an intervention, do we not reveal this very direction? To answer this, it is necessary to examine the features of an `intervention' on a quantum causal model. This can be done most easily by referring to what an intervention means for the corresponding functional model. To intervene on, say, the SIC-measurement $W$ in the circuit Fig. \ref{fig:intervene} (a), one would perform the following procedure: \\
I1. Discard the quantum system that would normally be input to the measurement $W$;\\
I2. Deterministically prepare a system in the desired state $\Pi_w$ and submit it as the output of $W$.\\
It is clear that this intervention leads to asymmetric statistics, since different choices of $w$ will have the power to influence the probabilities of the measurement outcomes of $D$, but not $A$. This is just to say that the intervention can signal into its causal future (as defined by the circuit) but not into its causal past. However, it would serve us well to examine the origin of this asymmetry to make sure that we have not sneaked it in by hand. On inspection, we see that we have: it was in the definition of the `intervention' where it was asserted that the `input' system should be discarded and the `output' re-prepared. On what basis are we allowed to assign these input/output labels prior to having information about the overall direction of causality? It seems that, prior to the intervention, the statistics $P(A,W,Z,D)$ would have been equally compatible with the QDAG Fig. \ref{fig:reverse} (a), whose functional model is now the time-reverse of the circuit in Fig. \ref{fig:intervene} (a). An intervention on \tit{this} circuit would produce the opposite effect, namely, it would permit signaling to $A$ but not to $D$, as shown in Fig. \ref{fig:reverse} (b). Thus, we cannot know \tit{a priori} the outcome of an intervention without knowing which of these QDAGs is the correct one, and this cannot be inferred from the statistics $P(A,W,Z,D)$. Thus, the very definition of an intervention \tit{requires} a pre-commitment to the input/output labelling and hence to the direction of causality, so we cannot appeal to interventions as a means to \tit{infer} this direction. Ultimately, experience shows that the correct choice is to align the causal direction with our own direction of time (i.e. the direction of entropy increase in the laboratory), but there is nothing in the causal model that can tell us this: we must put it in by hand.

\section{Conclusions}

At the end of the day, it is an incontrovertible fact that only one direction describes the direction of causality that we see in actual experiments. We have yet to perform an intervention on a variable that would allow us to influence the statistics of another variable in its causal past (and a Nobel Prize awaits the first person to do that!). However, in the context of the present discussion, this impossibility must not be seen as resulting from some preferred direction of causality intrinsic to the system, for we have just seen that there is no way to reveal such a preferred direction that does not end up putting it in by hand. Instead, we must conclude that the very real impossibility of sending information into our own past via interventions on physical systems is rather due to some physical constraint on the way that we -- as macroscopic physical systems -- are able to interact with the system. To the extent that the direction of causality is a determinate fact, therefore, it is determinate only relative to a choice of macroscopic observer. We may verify repeatedly that the causal relation $X \rightarrow Y$ holds in a system, but we have as yet no reason to exclude the possibility that some other observer interacting with the system could bring about the opposite relation. This possibility, at least, can be naturally accommodated in the symmetric causal model described here.

\appendix

\section{Appendix: Mathematical description of the model}

This Appendix lays out the bare mathematical structure of the proposed quantum causal model and the definitions for references in the main text. The following notation is used: uppercase roman letters such as $X$ denote random variables, and lowercase roman letters such as $x$ denote their possible values. Bold uppercase $\tbf{X}$ refers to a set of random variables, and bold lowercase $\tbf{x}$ to a corresponding set of possible values. Probability distributions and conditional distributions, like $P(\tbf{X}=\tbf{x}|Y=y) \, \forall \tbf{x},y$, will simply be abbreviated as $P(\tbf{X}|Y)$. The set union symbol `$\cup$' will often be omitted for the sake of tidy expressions. Note that although conditioning on `$Y$' is thus interpreted as conditioning on its \tit{value} (i.e. on $Y=y$), the similar-looking terms $\doo{Y},Y^{+},Y^{-}$ appearing below refer to pieces of information about $Y$ other than its value (indeed, in the case of $Y^{-}$, the value of $Y$ would not even be well-defined since $Y^{-}$ means that `$Y$ was not performed'). \\

\tbf{Definition: SIC-POVM}\\
A \tit{SIC-POVM} on a Hilbert space $\mathcal{H}_X$ of dimension $d_X$ is a set of $d^2_X$ linear operators $\{ \frac{1}{d} \Pi_{x} : x=1,2,...,d^2_X \}$ satisfying $\zum{x}{d^2_X} \, \frac{1}{d} \Pi_x = \mathbb{I}_X$ where $\mathbb{I}_X$ is the identity operator on $\mathcal{H}_X$, and also satisfying:
\eqn{
\tr{\Pi_x \Pi_{x'}} &=& \frac{d_X \delta_{x,x'}+1}{d_X+1} \, , \qquad x,x' \in \{1,2,...,d^2_X \} \, . \nonumber \\
&& 
}

\tbf{Definition: SIC-measurement} \\
A \tit{SIC-measurement} on a system with Hilbert space $\mathcal{H}_X$ is described by a \tit{SIC-POVM} $\{ \frac{1}{d} \Pi_x \}$. For a system in the state $\rho$ prior to measurement, the \tit{SIC-measurement} produces the outcome $x$ with probability $P(x)=\tr{\rho \, \frac{1}{d} \Pi_x }$. Conditional on the outcome $x$, the post-measurement state is $\Pi_x$.\\

\tbf{Definition: Possible explanation}\\
A causal graph $G(\tbf{X})$ is said to be a \tit{possible explanation} for a distribution $P(\tbf{X})$ on a set of $N$ variables $\tbf{X}$ if there exists a \tit{functional model} that has the same causal structure as $G(\tbf{X})$ and is capable of producing the statistics $P(\tbf{X})$ (see the next definition for details). For classical systems, the definition of a \tit{functional model} can be found in eg. Ref. \cite{PEARL,SGS}. For quantum systems, the following new definition is proposed:\\

\tbf{Definition: Quantum functional model}\\
A quantum \tit{functional model} for the variables $\tbf{X}$ is a quantum circuit having the following special properties:\\
(i) For every variable $X_i \in \tbf{X}$, $i=(1,2,...,N)$ the circuit contains a corresponding SIC-measurement whose outcomes correspond to the possible values $x_i$ of $X_i$ (Thus, a functional model only exists for variables with a square integer number of values);\\
(ii) Every input to the circuit is the maximally mixed state;\\
(iii) Aside from SIC-measurements, the only other processes in the circuit are unitary processes;\\
(iv) Every quantum output from the circuit is discarded (traced out). Only the classical data in the form of the joint probability $P(\tbf{X})$ of the outcomes of the SIC-measurements is retained.\\

The causal structure $G(\tbf{X})$ of a \tit{functional model} is then obtained by the following rule: $X_i$ is a parent of $X_j$ iff there is a wire in the circuit from the output of the measurement $X_i$ to the input of the measurement $X_j$ that is not intercepted by any other measurements in $\tbf{X}$ (though it may pass through unitary gates).\\

\tbf{Definition: Quantum Markov Condition}\\
Let $\tbf{U}$,$\tbf{V}$,$\tbf{W}$ be disjoint subsets of $\tbf{X}$. A pair $\{ P(\tbf{X}),G(\tbf{X}) \}$ is said to satisfy the \tit{Quantum Markov Condition} if the constraint $P(\tbf{U}\tbf{V}|\tbf{W})=P(\tbf{U}|\tbf{W})P(\tbf{V}|\tbf{W})$ holds whenever every undirected path between $\tbf{U}$ and $\tbf{V}$ is blocked by $\tbf{W}$. A path between two variables is said to be `blocked' by the set $\tbf{W}$ iff at least one of the following conditions holds:\\
\tbf{g-SSO:} There is a chain $A \rightarrow C \rightarrow B$ along the path whose middle member $C$ is in $\tbf{W}$;\\
\tbf{g-BK:} There is a collider $A \rightarrow C \leftarrow B$ on the path where $C$ is \tit{not} in $\tbf{W}$ and has no descendants in $\tbf{W}$.\\
\tbf{g-BK*:} There is a fork $A \leftarrow C \rightarrow B$ on the path where $C$ is \tit{not} in $\tbf{W}$ and has no ancestors in $\tbf{W}$.\\

\tbf{Claim:} Suppose that $G(\tbf{X})$ is a \tit{possible explanation} of $P(\tbf{X})$ (i.e. a \tit{functional model} exists with causal structure $G(\tbf{X})$ that produces $P(\tbf{X})$). Then the pair $\{ P(\tbf{X}),G(\tbf{X}) \}$ satisfies the \tit{Quantum Markov Condition} (This is argued in Ref. \cite{JACQUES}).\\

\tbf{Definition: QDAG}\\
A DAG $G(\tbf{X})$ is called a \tit{QDAG} if there exists a set of variables $\tbf{S}_{X_i}$ associated to every node $X_i \in \tbf{X}$  such that no member of $\{ \tbf{S}_{X_i} \cup X_i \}$ (called the \tit{slice} containing $X_i$) is a cause of any other member, and such that every path from an ancestor of $X_i$ to a descendant of $X_i$ is intercepted by at least one member of the slice, i.e. every path contains a causal chain $X \rightarrow Y \rightarrow Z$ whose middle member $Y$ is in $\{ \tbf{S}_{X_i} \cup X_i \}$.\\

\tbf{Definition: Quantum Causal Model}\\
A \tit{quantum causal model} is a pair $\{ P(\tbf{X}),G(\tbf{X}) \}$ that satisfies the \tit{Quantum Markov Condition}, where $G(\tbf{X})$ is a \tit{QDAG}, and for which the results of \tit{interventions} and \tit{un-measurements} are defined by the rules given in the definitions to follow (see Ref. \cite{JACQUES} for justification).\\ 

\tbf{Definition: Interventions}\\
Let $\{ P(\tbf{X}),G(\tbf{X}) \}$ be a quantum causal model. A measurement $Y$ introduced into this model is called an \tit{intervention} on some variable $W \in \tbf{X}$ iff the resulting model, denoted $\{ P(\tbf{X},Y|\doo{Y}), G(\tbf{X},Y|\doo{Y}) \}$, satisfies the following properties:\\

\noindent (i) $Y$ is a direct cause of $W$ and only $W$ in $G(\tbf{X},Y|\doo{Y})$;\\
(ii) The performance or non-performance of the intervention $Y$ on $W$ does not affect the conditional distribution of variables that are not descendants of $W$. Formally, let $\tbf{A},\tbf{R}$ denote the non-descendants of $W$, and let $\doo{Y}$ represent the information `Y was performed' and $Y^-$ the information that `Y was not performed'. Then:
\eqn{
P(\tbf{A},\tbf{R}|W,\doo{Y}) = P(\tbf{A},\tbf{R}|W,Y^{-}) \, .
}
(iii) The performance or non-performance of the intervention $Y$ on $W$ does not affect the conditional distribution of variables that are not ancestors of $W$. Formally, let $\tbf{D},\tbf{R}$ denote the non-ancestors of $W$. Then:
\eqn{
P(\tbf{D},\tbf{R}|W,\doo{Y}) = P(\tbf{D},\tbf{R}|W,Y^{-}) \, .
}
(iv) The value of $W$ is equal to the value of $Y$, hence $W$ has no other causes in $G(\tbf{X},Y|\doo{Y})$;\\
(v) $Y$ has no parents in the system, i.e. it is exogenous in $G(\tbf{X},Y|\doo{Y}) $.\\

When an intervention $Y$ is performed on $W$ in the causal model $\{ P(\tbf{X}),G(\tbf{X}) \}$, the resulting statistics  $P(\tbf{X},Y|\doo{Y})$ are given by:
\eqn{
P(\tbf{X}|Y, \doo{Y}) =&& \nonumber \\
 P(\tbf{D},\tbf{R}_D|W,\tbf{S}_W,Y^{-}) \, && \delta(W,Y) \, P(\tbf{S}_W, \tbf{A},\tbf{R}_A|W,Y^{-})  \, , \nonumber \\ 
&&
}
where $\{ \tbf{S}_W \cup W \}$ is the slice containing $W$, and $\tbf{R}_D$ (resp. $\tbf{R}_A$) are the descendants (resp. ancestors) of $\tbf{S}_W$ that are not descendants (resp. ancestors) of $W$ in $G(\tbf{X})$. The DAG $G(\tbf{X},Y|\doo{Y})$ resulting from the intervention is obtained by the same procedure as in the classical case, i.e. by deleting the parents of $W$ and introducing $Y$ as its sole parent.\\

\tbf{Definition: Un-measurements}\\
Consider a quantum causal model $\{ P(\tbf{X},Z),G(\tbf{X},Z) \}$ and let $\tbf{D}$ (resp. \tbf{A}) be the descendents (resp. ancestors) of $Z$ in $G(\tbf{X},Z)$ and let $\tbf{R}$ be the remaining variables excluding $Z$. Then the statistics that one would obtain if the measurement $Z$ were not to be performed (i.e. the result of \tit{un-measuring} $Z$) are given by:
\eqn{ \label{eqn:unmeas}
P(\tbf{X}|\,Z^{-}) &=& \nonumber \\
\zum{z}{d^2_Z} \, P(\tbf{D}|\tbf{A},\tbf{R},z,\,&Z^{+}&) \left[ (1+d_Z) P(z|\tbf{A},\tbf{R},\, Z^{+})-\frac{1}{d_Z} \right] \nonumber \\ && \times P(\tbf{A}, \tbf{R}|\,Z^{+}) \, , \nonumber \\
&&
}
where $Z^{+}$ denotes the information that `$Z$ was performed' and $Z^{-}$ denotes that `$Z$ was not performed'. The DAG $G(\tbf{X})$ resulting from the \tit{un-measurement} of $Z$ is obtained from $G(\tbf{X},Z)$ by deleting $Z$ and its connected arrows, and then connecting every former parent of $Z$ to every former child of $Z$.\\

\end{document}